\documentclass[aps,prd,preprint,groupedaddress]{revtex4}

\usepackage{epsfig}

\begin{document}

\def\d{{\rm d}}
\def\hs{\theta_{\tilde{q}}}
\def\hp{\theta_{\tilde{t}}}
\def\hm{\theta_{\tilde{b}}}
\def\mg{m_{\tilde{g}}}
\def\ms{m_{\tilde{q}}}
\def\mi{m_{\tilde{\chi}^0_i}}
\def\mj{m_{\tilde{\chi}^0_j}}
\def\O{{\cal O}}
\def\p{I\!\!P}
\def\tg{t_{\tilde{g}}}
\def\ti{t_{\tilde{\chi}^0_i}}
\def\tj{t_{\tilde{\chi}^0_j}}

\def\lp{\left. }
\def\rp{\right. }
\def\lr{\left( }
\def\rr{\right) }
\def\le{\left[ }
\def\re{\right] }
\def\lg{\left\{ }
\def\rg{\right\} }
\def\lb{\left| }
\def\rb{\right| }

\def\beq{\begin{equation}}
\def\eeq{\end{equation}}
\def\bea{\begin{eqnarray}}
\def\eea{\end{eqnarray}}

\preprint{hep-ph/0507073}
\preprint{LPSC 05-048}
\title{Non-Diagonal and Mixed Squark Production at Hadron Colliders}
\author{Giuseppe Bozzi, Benjamin Fuks and Michael Klasen}
\email[]{klasen@lpsc.in2p3.fr}
\affiliation{Laboratoire de Physique Subatomique et de Cosmologie,
 Universit\'e Joseph Fourier/CNRS-IN2P3, 53 Avenue des Martyrs,
 F-38026 Grenoble, France}
\date{\today}
\begin{abstract}
We calculate squared helicity amplitudes for non-diagonal and mixed squark
pair production at hadron colliders, taking into account not only
loop-induced QCD diagrams, but also previously unconsidered electroweak
channels, which turn out to be dominant. Mixing effects are included for
both top and bottom squarks. Numerical results are presented for several
SUSY benchmark scenarios at both the CERN LHC and the Fermilab Tevatron,
including the possibilities of light stops or sbottoms. The latter should be
easily observed at the Tevatron in associated production of stops and
sbottoms for a large range of stop masses and almost independently of the
stop mixing angle. Asymmetry measurements for light stops at the polarized
BNL RHIC collider are also briefly discussed.
\end{abstract}
\maketitle


\section{Introduction}
\label{sec:1}

One of the most promising extensions of the Standard Model (SM) of particle
physics is the Minimal Supersymmetric Standard Model (MSSM)
\cite{Nilles:1983ge,Haber:1984rc}, which postulates a symmetry between
fermionic and bosonic degrees of freedom in nature and predicts the
existence of a fermionic (bosonic) supersymmetric (SUSY) partner for each
bosonic (fermionic) SM particle. Since SUSY and SM particles contribute to
the quadratic divergence of the mass of the Higgs boson with equal strength,
but opposite sign, the MSSM can, {\it inter alia}, stabilize the electroweak
energy scale with respect to the Planck scale and thus propose a solution to
the hierarchy problem.

Unfortunately, SUSY particles still remain to be discovered. Their masses
must therefore be considerably larger than those of the corresponding SM
particles, and the symmetry is bound to be broken. In order to remain a
viable solution to the hierarchy problem, SUSY can, however, only be broken
via soft mass terms in the Lagrangian, with the consequence that the SUSY
particle masses should lie in the TeV range and thus within the discovery
reach of current and future hadron colliders such as the Tevatron and the
LHC.

Due to their strong coupling, squarks should be abundantly produced at
hadron colliders. In addition, phase space favors the production of the
lighter of the two squark mass eigenstates of identical flavor, which are
superpositions of the left- and right-handed helicity eigenstates. Since the
off-diagonal elements of the two-dimensional squark mixing matrix are
proportional to the mass of the corresponding SM quark, squark mixing is
particularly important for third generation squarks. As a consequence,
hadroproduction of light stop \cite{Beenakker:1997ut} and, more recently,
also light sbottom pairs \cite{Berger:2000mp} has received particular
theoretical and experimental attention. Mixing effects have also been
analyzed recently for slepton, and in particular stau, hadroproduction
\cite{Bozzi:2004qq}.

As has been observed previously, Quantum Chromodynamics (QCD) alone does not
allow for non-diagonal (light plus heavy) squark hadroproduction at
tree-level, {\it i.e.} ${\cal O}(\alpha_s^2)$ in the strong coupling
constant $\alpha_s$. Non-diagonal squark production rather requires the
presence of at least one squark-mixing vertex, such as the four-squark
interaction, present, {\it e.g.}, in final-state rescattering of
gluon-induced diagonal squark production. The corresponding finite one-loop
diagrams at ${\cal O}(\alpha_s^4)$ have been evaluated in a decoupled gluino
scenario with top squark loops only and found to be suppressed with respect
to diagonal light/heavy stop pair production at the Tevatron/LHC by three to
six orders of magnitude \cite{Beenakker:1997ut}.

In this Paper, we investigate the importance of electroweak channels for
non-diagonal and mixed squark pair production at hadron colliders. Naively,
one expects these cross sections, which are of ${\cal O}(\alpha^2)$ in the
fine structure constant $\alpha$, to be smaller than the diagonal strong
channels by about two orders of magnitude. For non-diagonal squark
production, the interplay between loop suppression in QCD and coupling
suppression in the electroweak case merits a detailed investigation. In the
presence of the mixing of bottom squarks, their loop contributions must also
be taken into account. Mixed top and bottom squark production is possible at
tree-level only through an $s$-channel exchange of a charged $W^\pm$-boson.
Observation of this channel may allow for interesting conclusions on the
supersymmetric version of the CKM matrix.

Our analytical calculations are presented in Sec.\ \ref{sec:2}, where we
put in the additional effort to calculate squared amplitudes for definite
initial parton helicities to exhibit clearly the mixing angle dependence
in the left- and right-handed components of the electroweak currents and to
allow for future applications to polarized hadron collisions. To be
complete, we present our results for diagonal, non-diagonal, and mixed top
and bottom squark hadroproduction with quark and gluon initial states up to
${\cal O}(\alpha_s^4)$ and ${\cal O}(\alpha^2)$, respectively.

In Sec.\ \ref{sec:3}, we choose several typical SUSY mass spectra arising
from different SUSY breaking scenarios, including those allowing for light
stops and sbottoms, and apply them to the Tevatron and the LHC. Cross
sections are presented as functions of the squark masses, mixing angles, and
general SUSY breaking mass parameters. We also briefly discuss possible
asymmetry measurements for light stops at the polarized RHIC collider. Our
conclusions are given in Sec.\ \ref{sec:4}, and squark mixing is discussed
in App.\ \ref{sec:a}.

It is not the aim of this work to present a full signal-to-background
analysis of non-diagonal and mixed squark production, as this is best done
within the experimental collaborations and using full detector simulations.
Rather, our analytical results lend themselves easily to implementation in
general purpose Monte Carlo programs such as PYTHIA \cite{Sjostrand:2003wg}
or HERWIG \cite{Corcella:2002jc}, which are traditionally employed for these
kinds of simulations.

\section{Analytical results}
\label{sec:2}

In the following, we present the leading contributions in the strong
($\alpha_s$) and electromagnetic ($\alpha$) coupling constant to the
color-averaged cross sections $\d\hat{\sigma}_{h_a,h_b}$ for definite
helicities $h_a$ and $h_b$ of the initial partons $a$ and $b$.
We define the square of the weak coupling constant $g_W^2=e^2/\sin^2
\theta_W$ in terms of the electromagnetic fine structure constant
$\alpha=e^2/(4\pi)$ and the squared sine of the electroweak mixing angle
$x_W=\sin^2\theta_W$. The coupling strengths of left- and right-handed
(s)quarks to the neutral electroweak current are then given by
\bea
 L_q = 2\,T^{3}_q-2\,e_q\,x_W &\mbox{~~and~~}&
 R_q =           -2\,e_q\,x_W,
\eea
where the weak isospin quantum numbers are $T_q^3=\pm1/2$ for left-handed
and $T_q^3=0$ for right-handed up- and down-type (s)quarks, and their
fractional electromagnetic charges are denoted by $e_q$.

In general SUSY breaking models, where the squark interaction eigenstates
are not identical to the respective mass eigenstates, the coupling strengths
$L_q$ and $R_q$ must be multiplied by $S_{j1}S_{i1}^\ast$ and $S_{j2}
S_{i2}^\ast$, respectively, where $i,j\in\{1,2\}$ label the squark mass
eigenstates (conventionally $m_{\tilde{q}_1} < m_{\tilde{q}_2}$) and $S$
represents the unitary matrix diagonalizing the squark mass matrix
(see App.\ \ref{sec:a}). In the following, the dependence on the squark
mixing angle $\theta_{\tilde{q}}$ will, however, be presented explicitly.

Our results for the strong or electroweak $2\to2$ scattering processes
\bea
 a_{h_a}(p_a)b_{h_b}(p_b)\to\tilde{q}_i(p_1)\tilde{q}^{(\prime)*}_j(p_2)
\eea
with $a,b=q,\bar{q},g$ will be expressed in terms of the squark masses
$m_{\tilde{q}^{(\prime)}_{i,j}}$, the conventional Mandelstam variables,
\bea
 s=(p_a+p_b)^2 &~~,~~& t=(p_a-p_1)^2 \mbox{~~,~~and~~} u=(p_a-p_2)^2,
\eea
and the masses of the neutral and charged electroweak gauge bosons $m_Z$
and $m_W$. Unpolarized cross sections, averaged over initial spins, can then
easily be derived from the expression
\bea
 \d\hat{\sigma}&=&
 {\d\hat{\sigma}_{ 1, 1}
 +\d\hat{\sigma}_{ 1,-1}
 +\d\hat{\sigma}_{-1, 1}
 +\d\hat{\sigma}_{-1,-1}
 \over 4},
 \label{eq:4}
\eea
while single and double polarized cross sections, including the same
average factor for initial spins, are given by
\bea
 \d\Delta\hat{\sigma}_{L~~}&=&
 {\d\hat{\sigma}_{ 1, 1}
 +\d\hat{\sigma}_{ 1,-1}
 -\d\hat{\sigma}_{-1, 1}
 -\d\hat{\sigma}_{-1,-1}
 \over 4}
\eea
and
\bea
 \d\Delta\hat{\sigma}_{LL }&=&
 {\d\hat{\sigma}_{ 1, 1}
 -\d\hat{\sigma}_{ 1,-1}
 -\d\hat{\sigma}_{-1, 1}
 +\d\hat{\sigma}_{-1,-1}
 \over 4},
\eea
so that the the single and double longitudinal spin asymmetries become
\bea
 A_{L } ~=~ {\d\Delta\hat{\sigma}_{L~}\over\d\hat{\sigma}} &{\rm and}&
 A_{LL} ~=~ {\d\Delta\hat{\sigma}_{LL}\over\d\hat{\sigma}}.
\eea

\subsection{Diagonal top and bottom squark production}

\begin{figure}
 \centering
 \includegraphics[width=0.70\columnwidth]{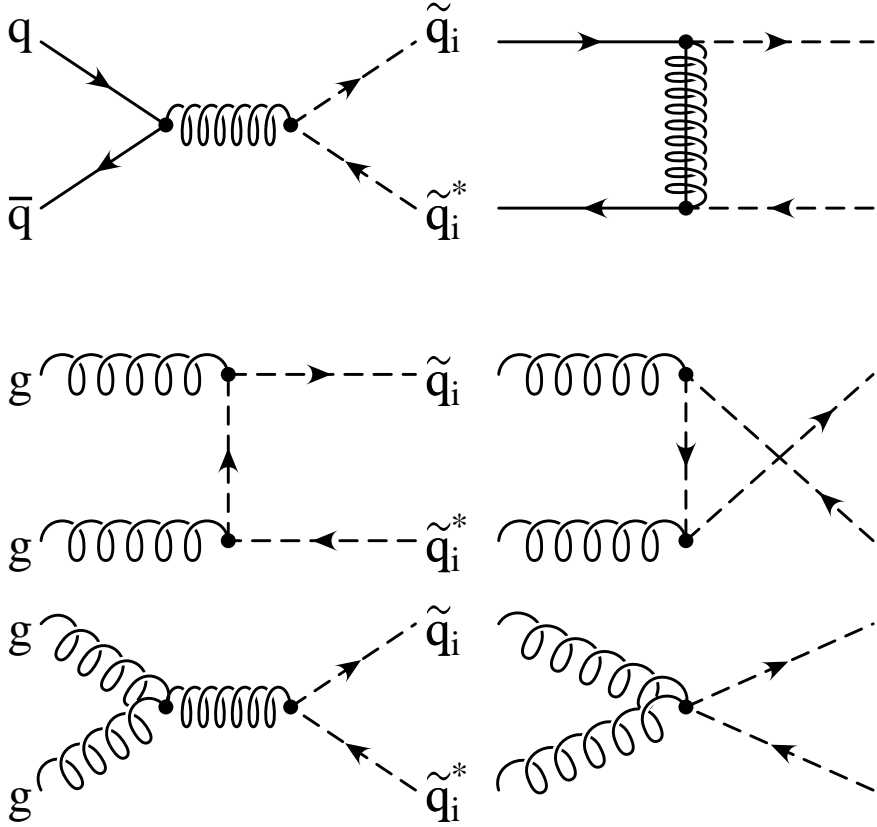}
 \caption{\label{fig:1}Tree-level QCD Feynman diagrams for diagonal squark
 production at hadron colliders. The upper right diagram is absent for top
 squarks due to the negligible top quark density in the proton.}
\end{figure}

For diagonal top and bottom squark production, we consider only the
dominant tree-level strong coupling channels shown in Fig.\ \ref{fig:1}.
%
%
For next-to-leading order QCD corrections to diagonal top squark production
see Ref.\ \cite{Beenakker:1997ut}. Other exchanges are significantly
suppressed by smaller electroweak or Yukawa couplings. For the two upper
diagrams, initiated by quarks, we find
\bea
 {\d\hat{\sigma}_{h_a,h_b}^{q\bar{q}}\over\d t}
 &=& {4\pi\alpha_s^2\over 9s^2}\,\le1-h_ah_b\re\,{tu-\ms^4\over s^2}
     \nonumber\\
 &-& {4\pi\alpha_s^2\over27s^2}\,\le1\pm(h_b-h_a)\cos2\hs-h_ah_b\re\,
     {tu-\ms^4\over s\tg}\,\delta_{qb}\,\delta_{\tilde{q}\tilde{b}}
     \nonumber\\
 &+& { \pi\alpha_s^2\over18s^2}\ \le
     {(1+h_a h_b)\,(1-\cos4\hs)\,\mg^2 s+(1-h_a h_b)\,(3+\cos4\hs)\,
      (t u-\ms^4)\over\tg^2} \rp \nonumber\\
 & & \lp \hspace*{10mm}\pm\,4 \cos2\hs\,(h_b-h_a){tu-\ms^4\over\tg^2}\re
     \delta_{qb}\,\delta_{\tilde{q}\tilde{b}},
\eea
where the upper sign holds for $\tilde{b}_1$ and the lower sign for
$\tilde{b}_2$ production. Stops are produced only through $s$-channel
gluon exchange due to negligible top quark parton density functions (PDFs)
in the proton. Even for sbottom production, $t$-channel gluino contributions
are suppressed by small bottom PDFs and the heavy gluino propagator,
$t_{\tilde{g}}^{-1}=(t-\mg^2)^{-1}$, where $\mg$ is the gluino mass. In the
case of no squark mixing, our results agree with the double-polarized and
unpolarized cross sections in Ref.\ \cite{Gehrmann:2004xu}.

For the four lower diagrams in Fig.\ \ref{fig:1}, initiated by gluons, we
find
\bea
 {\d\hat{\sigma}_{h_a,h_b}^{gg}\over \d t}
 &=& {\pi\alpha_s^2\over 128s^2}
 \le 24 \lr 1-2{t_{\tilde{q}} u_{\tilde{q}}\over s^2}\rr - {8\over3}\re
 \le (1-h_a h_b)-2 {s \ms^2 \over t_{\tilde{q}} u_{\tilde{q}}}
 \lr (1-h_a h_b)-  {s \ms^2 \over t_{\tilde{q}} u_{\tilde{q}}}\rr\re,
\eea
where $t_{\tilde{q}}=t-m_{\tilde{q}}^2$ and $u_{\tilde{q}}=u-m_{\tilde{q}}
^2$, independently of the squark mixing angle and thus in direct agreement
with the double-polarized and unpolarized cross sections of Ref.\
\cite{Gehrmann:2004xu}.

\subsection{Non-diagonal top and bottom squark production}

\begin{figure}
 \centering
 \includegraphics[width=0.35\columnwidth]{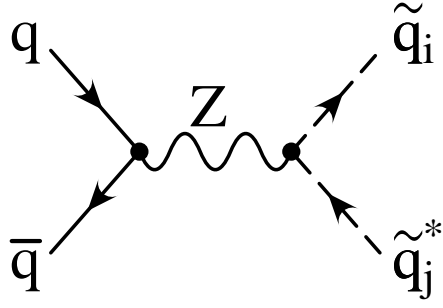}
 \caption{\label{fig:2}Dominant tree-level electroweak Feynman diagram for
 non-diagonal ($i\neq j$) squark production at hadron colliders.}
\end{figure}

\begin{figure}
 \centering
 \includegraphics[width=0.70\columnwidth]{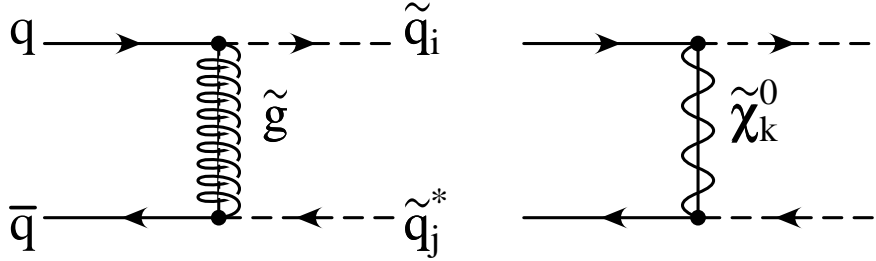}
 \caption{\label{fig:3}Feynman diagrams for non-diagonal ($i\neq j$)
 squark production at hadron colliders proceeding through $t$-channel
 gluino (left) and neutralino (right) exchange.}
\end{figure}

Due to the non-chiral strong interaction of the gluon, non-diagonal squark
pairs cannot be produced in QCD at tree-level with either $q\bar{q}$ or
$gg$ initial states. There are, however, quark-induced chiral tree-level
electroweak and supersymmetric interactions, proceeding either through an
$s$-channel $Z$-boson exchange (see Fig.\ \ref{fig:2}) or a $t$-channel
gluino or
%
%
neutralino exchange (see Fig.\ \ref{fig:3}). As in the case of diagonal
%
%
squark production, $t$-channel exchanges are present only for bottom
squarks, due to negligible top quark PDFs, and even there they are
suppressed by small bottom PDFs and the heavy gluino or neutralino
propagators, where the latter are defined by $t_{\tilde{\chi}_k^0}^{-1}=
(t-m_{\tilde{\chi}_k^0}^2)^{-1}$. $m_{\tilde{\chi}_k^0}$ is the mass of the
$k^{\rm th}$ neutralino mass eigenstate, which couples with strength
\bea
 \lg L_{\tilde{q}i},R_{\tilde{q}i}^\ast \rg & = & \pm \le
 e_q \sqrt{x_W(1-x_W)} N_{i1}' + 
 \lr T^{3}_{q_{L,R}} - e_q x_W \rr N_{i2}' \re
\eea
to left- and right-handed (s)quarks, if we neglect the relatively small
bottom Yukawa couplings. The unitary matrix $N'$ diagonalizes the
neutralino mass matrix in the photino-zino basis \cite{Gunion:1984yn}.
Gluino exchanges are, in contrast, enhanced by the strong coupling
constant $\alpha_s$.

Our analytical result up to ${\cal O}(\alpha^2)$ and ${\cal O}(\alpha_s^2)$
for $s$- and $t$-channel exchanges, including interferences, is
\bea
 {\d\hat{\sigma}_{h_a,h_b}^{q\bar{q}}\over\d t}
 &=& \ {\pi\alpha^2\over s^2}\ 
     \le{tu-m_{\tilde{q}_1}^2 m_{\tilde{q}_2}^2\over s^2}\re
     {\sin^2(2\hs)\le(1-h_a)(1+h_b)L_q^2+(1+h_a)(1-h_b)R_q^2\re\over
      32 x_W^2 (1-x_W)^2(1-m_Z^2/s)^2} \nonumber \\
 &-& {\pi\alpha\alpha_s\over s^2}
     \le{tu-m_{\tilde{q}_1}^2 m_{\tilde{q}_2}^2\over s\tg}\re
     {\sin^2(2\hs)\le(1-h_a)(1+h_b)L_q-(1+h_a)(1-h_b)R_q\re\over
      9 x_W (1-x_W)(1-m_Z^2/s)}\
     \delta_{qb}\,\delta_{\tilde{q}\tilde{b}} \nonumber \\
 &+& \ {\pi\alpha_s^2\over9s^2}\ 
     \le{\le(1+h_ah_b)(3+\cos4\hs)-4(h_a+h_b)\cos2\hs\re\mg^2 s\over\tg^2}
     \rp \nonumber \\
 & & \lp \hspace*{10mm} + {2\sin^22\hs(1-h_ah_b)
     (tu-m_{\tilde{q}_1}^2 m_{\tilde{q}_2}^2)\over \tg^2}\re
     \delta_{qb}\,\delta_{\tilde{q}\tilde{b}} \nonumber \\
 &-& {\pi\alpha^2\over s^2}\sum_i
     \delta_{qb}\,\delta_{\tilde{q}\tilde{b}}
     \le{tu-m_{\tilde{q}_1}^2 m_{\tilde{q}_2}^2\over s\ti}\re
     {\sin^2(2\hs)\le(1-h_a)(1+h_b)L_qL_{\tilde{q}i}^2-(1+h_a)(1-h_b)
     R_qR_{\tilde{q}i}^2\re\over 12 x_W^2 (1-x_W)^2 (1-m_Z^2/s)}\nonumber \\
 &+& {\pi\alpha^2\over s^2}\sum_{i,j}
     {\delta_{qb}\,\delta_{\tilde{q}\tilde{b}}\over1+\delta_{ij}}
     \le{\le(1+h_ah_b)(3+\cos4\hs)-4(h_a+h_b)\cos2\hs\re
     L_{\tilde{q}i} L_{\tilde{q}j} R_{\tilde{q}i} R_{\tilde{q}j}\,
     \mi \mj s\over x_W^2 (1-x_W)^2 \ti\tj}\rp \nonumber \\
 & & \lp \hspace*{10mm} + {\sin^22\hs\le(1-h_a)(1+h_b)L_{\tilde{q}i}^2
     L_{\tilde{q}j}^2+(1+h_a)(1-h_b)R_{\tilde{q}i}^2R_{\tilde{q}j}^2\re
     (tu-m_{\tilde{q}_1}^2 m_{\tilde{q}_2}^2)\over x_W^2(1-x_W)^2\ti\tj}\re,
 \label{eq:7}
\eea
where we have summed over $\tilde{q}_1\tilde{q}_2^\ast+\tilde{q}_2
\tilde{q}_1^\ast$ final states. Note that there is no neutralino-gluino
interference term due to color (non-)conservation.

\begin{figure}
 \centering
 \includegraphics[width=\columnwidth]{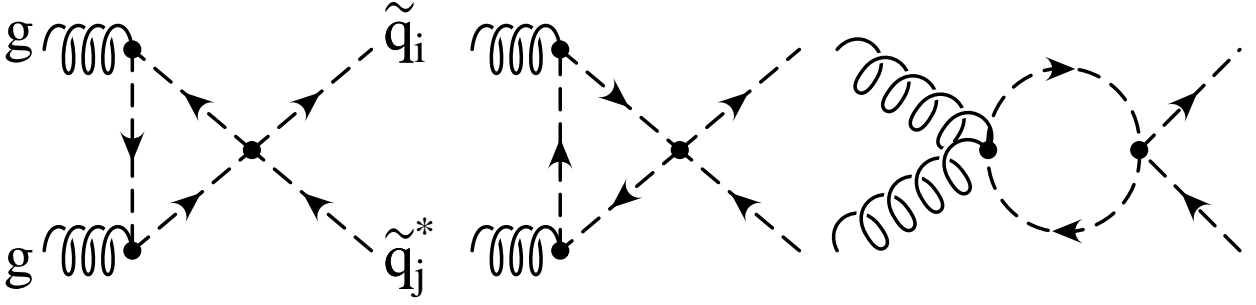}
 \caption{\label{fig:4}Subdominant loop-level QCD Feynman diagrams for
 non-diagonal ($i\neq j$) squark production at hadron colliders.}
\end{figure}

Within QCD and in the limit of a decoupled gluino, the only possibility to
produce non-diagonal squark pairs is by rescattering of diagonal squark
pairs through four-squark vertices in the final state. The corresponding
Feynman diagrams, shown in Fig.\ \ref{fig:4},
%
%
have (one-)loop topology and are therefore suppressed by additional squark
propagators and/or annihilations within the squark loop. The squared
helicity amplitude for the production of non-diagonal stop pairs in
gluon-gluon collisions is given by
\begin{eqnarray}
\frac{d\hat{\sigma}_{h_a,h_b}^{gg}}{dt} &=&
 (1+h_a h_b) \le \frac{37 \alpha_{s}^{4}
\sin^2{\left(4 \theta_{\tilde{t}}\right)}}{27648 \pi
s^{4}} |\Delta \ln_{\tilde{t}}|^2\rp \nonumber\\
&+& \sum_{\tilde{q} \neq \tilde{t}} \frac{5 \alpha_{s}^{4}
\cos^2{\left(2 \theta_{\tilde{q}}\right)} \sin^2{\left(2
\theta_{\tilde{t}}\right)}}{3072 \pi
s^{4}} |\Delta \ln_{\tilde{q}}|^2\nonumber\\
&+& \sum_{\tilde{q} \neq \tilde{t}} \frac{5 \alpha_{s}^{4}
\cos{\left(2 \theta_{\tilde{q}}\right)} \cos{\left(2
\theta_{\tilde{t}}\right)} \sin^2{\left(2
\theta_{\tilde{t}}\right)}}{2304 \pi s^{4}}
{\rm Re}\left(\Delta \ln_{\tilde{q}} \Delta
\ln_{\tilde{t}}\right)\nonumber\\
&+& \lp \sum_{\tilde{q},\tilde{q}'\neq \tilde{t}; \tilde{q}\neq
\tilde{q}'} \frac{5 \alpha_{s}^{4} \cos{\left(2
\theta_{\tilde{q}}\right)} \cos{\left(2
\theta_{\tilde{q}'}\right)} \sin^2{\left(2
\theta_{t}\right)}}{1536 \pi s^{4}} {\rm Re}\left(\Delta
\ln_{\tilde{q}} \Delta \ln_{\tilde{q'}}\right)\re,
\end{eqnarray}
where $\Delta \ln_{\tilde{q}} = m_{\tilde{q}_1}^2
\ln^2{\left(-x_{\tilde{q}_1}\right)} - m_{\tilde{q}_2}^2
\ln^2{\left(-x_{\tilde{q}_2}\right)}$,
\bea
x_{\tilde{q}_{i}}&=&\frac{1-\beta_{\tilde{q}_{i}}}{1+\beta_{\tilde{q_{i}}}},
\eea
and $\beta_{\tilde{q}_i}=\sqrt{1-4m_{\tilde{q}_i}^2}$ is the velocity of the
$i^{\rm th}$ squark mass eigenstate. In the limit of degenerate light
squarks, only top and bottom squark loops survive loop annihilations, and
the squared helicity amplitude simplifies to
\begin{eqnarray}
\frac{d\hat{\sigma}_{h_a,h_b}^{gg}}{dt} &=&
  (1+h_a h_b) \le  \frac{37 \alpha_{s}^{4}
\sin^2{\left(4 \theta_{\tilde{t}}\right)}}{27648 \pi
s^{4}} |\Delta \ln_{\tilde{t}}|^2\rp\nonumber\\
&+& \frac{5 \alpha_{s}^{4} \cos^2{\left(2
\theta_{\tilde{b}}\right)} \sin^2{\left(2
\theta_{\tilde{t}}\right)}}{3072 \pi
s^{4}} |\Delta \ln_{\tilde{b}}|^2\nonumber\\
&+& \lp \frac{5 \alpha_{s}^{4} \cos{\left(2 \theta_{\tilde{b}}\right)}
\cos{\left(2 \theta_{\tilde{t}}\right)} \sin^2{\left(2
\theta_{\tilde{t}}\right)}}{2304 \pi s^{4}}
{\rm Re}\left(\Delta \ln_{\tilde{b}} \Delta
\ln_{\tilde{t}}\right)\re.
\label{eq:14}
\end{eqnarray}
These expressions have been summed over $\tilde{t}_1\tilde{t}_2^\ast+
\tilde{t}_2\tilde{t}_1^\ast$ final states and generalize the corresponding
result in Ref.\ \cite{Beenakker:1997ut}, where only top squark loops were
taken into account. For non-diagonal sbottom production, top and bottom
squark indices have to be exchanged in the equations above.

\subsection{Mixed top and bottom squark production}

\begin{figure}
 \centering
 \includegraphics[width=0.35\columnwidth]{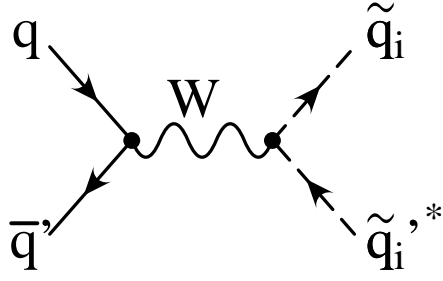}
 \caption{\label{fig:5}Dominant electroweak Feynman diagram for mixed
 ({\it e.g.} top and bottom) squark production at hadron colliders.}
\end{figure}

As mentioned above, mixed top and bottom squark production proceeds at
tree-level only through an $s$-channel exchange of a charged $W^\pm$-boson,
shown in Fig.\ \ref{fig:5}. The exchange of a $t$-channel chargino is
%
%
excluded by negligible top quark PDFs in the proton, and $t$-channel
gluino contributions are loop-suppressed, since they require electroweak
rescattering of the intermediate squarks. The squared helicity amplitude
is therefore
\bea
 {\d\hat{\sigma}_{h_a,h_b}^{q\bar{q}'}\over\d t}
 &=& \ {\pi\alpha^2\over s^2}\ |V_{qq'}|^2 |V_{\tilde{t}_1\tilde{b}_1}|^2
     \le{tu-m_{\tilde{t}_1}^2 m_{\tilde{b}_1}^2\over s^2}\re
     {\cos^2\hp\,\cos^2\hm\,(1-h_a)(1+h_b)\over
      4 x_W^2 (1-m_W^2/s)^2},
\eea
where we show explicitly the dependence on the SM and SUSY CKM matrix
elements. Here, we have {\em not} summed over $\tilde{t}_1\tilde{b}_1^\ast+
\tilde{b}_1\tilde{t}_1^\ast$ final states, since each of them is produced
by different partonic luminosities of the correct weak isospin partners $q$
and $\bar{q}'$ in initial state. This purely left-handed charged current
cross section is easily derived from the squared $s$-channel contribution in
Eq.\ (\ref{eq:7}) by adjusting the squark masses, including the squared
absolute values of the appropriate CKM matrix elements $V_{qq'}$ and
$V_{\tilde{t}_1\tilde{b}_1}$ and by setting
\bea
 m_Z~\rightarrow~m_W,~~~
 R_q=0,~~~
 L_q=\sqrt{2}\cos^2\theta_W, &\mbox{~~~and~~~}&
 \sin^2(2\hs)~\rightarrow~4\cos^2\hp\cos^2\hm.
\eea
For the mixed production of the heavier squark mass eigenstates, the
corresponding index 1 has to be replaced by 2 and the squared cosine of the
mixing angle by the squared sine. At lowest order, gluon initial states do
not allow for the production of a (charged) mixed squark final state.

\section{Numerical results}
\label{sec:3}

We now turn to our numerical results for the production of diagonal,
non-diagonal, and mixed squark pairs at hadron colliders and use the squared
helicity amplitudes given in Sec.\ \ref{sec:2} together with Eq.\
(\ref{eq:4}) to calculate unpolarized partonic cross sections. The QCD
factorization theorem then allows to compute unpolarized hadronic cross
sections
\bea
 \sigma &~=&
 \int_{m^2/S}^1\!\d\tau\!
 \int_{-1/2\ln\tau}^{1/2\ln\tau}\!\!\d y
 \int_{t_{\min}}^{t_{\max}} \d t \
 f_{a/A}(x_a,M_a^2) \ f_{b/B}(x_b,M_b^2) \ {\d\hat{\sigma}\over\d t}
\eea
by convolving the relevant partonic cross section d$\hat{\sigma}$ with
universal parton densities $f_{a/A}$ and $f_{b/B}$ of partons $a,b$ in the
hadrons $A,B$. The PDFs depend on the longitudinal momentum fractions of the
two partons $x_{a,b} = \sqrt{\tau}e^{\pm y}$ and on the unphysical
factorization scales $M_{a,b}$. We use the most recent leading order (LO)
PDF set by the CTEQ collaboration, CTEQ6L1 \cite{Pumplin:2002vw}, at the
factorization scale $M_a=M_b=m=(m_{\tilde{q}_i}+m_{\tilde{q}_j^{(\prime)}})
/2$ and identify the latter with the renormalization scale $\mu$ in the
strong coupling constant $\alpha_s(\mu)$. The QCD scale parameter $\Lambda$
in the CTEQ6L1 fit for $n_f=5$ active quark flavors is 165 MeV.

For the masses and widths of the electroweak gauge bosons, we use the
current values of $m_Z=91.1876$ GeV, $m_W=80.425$ GeV, $\Gamma_Z=2.4952$
GeV, and $\Gamma_W=2.124$ GeV. The squared sine of the electroweak mixing
angle
\bea
 \sin^2\theta_W &=& 1-m_W^2/m_Z^2
\eea
and the electromagnetic fine structure constant
\bea
 \alpha &=& \sqrt{2} G_F m_W^2 \sin^2\theta_W / \pi
\eea
can then be calculated in the improved Born approximation using the world
average value of $G_F=1.16637\cdot 10^{-5}$ GeV$^{-2}$ for Fermi's coupling
constant \cite{Eidelman:2004wy}. For the SM CKM matrix elements, we take the
central values in Ref.\ \cite{Eidelman:2004wy}, while we set the (so far
unknown) SUSY CKM matrix elements $V_{\tilde{t}_i \tilde{b}_j}$ to one.

The physical masses of the produced squark mass eigenstates and mixing
angles are calculated using the recently updated computer program SUSPECT
\cite{Djouadi:2002ze}. Its Version 2.3 includes now a consistent calculation
of the Higgs mass, with all one-loop and the dominant two-loop radiative
corrections, in the renormalization group equations, that link the
restricted set of SUSY breaking parameters at the gauge coupling unification
scale to the complete set of observable SUSY masses and mixing angles at the
electroweak scale. We choose two recently proposed minimal supergravity
(mSUGRA) points, SPS 1a and SPS 5, as benchmarks for our numerical study
\cite{Allanach:2002nj}.

\begin{figure}
 \centering
 \includegraphics[width=\columnwidth]{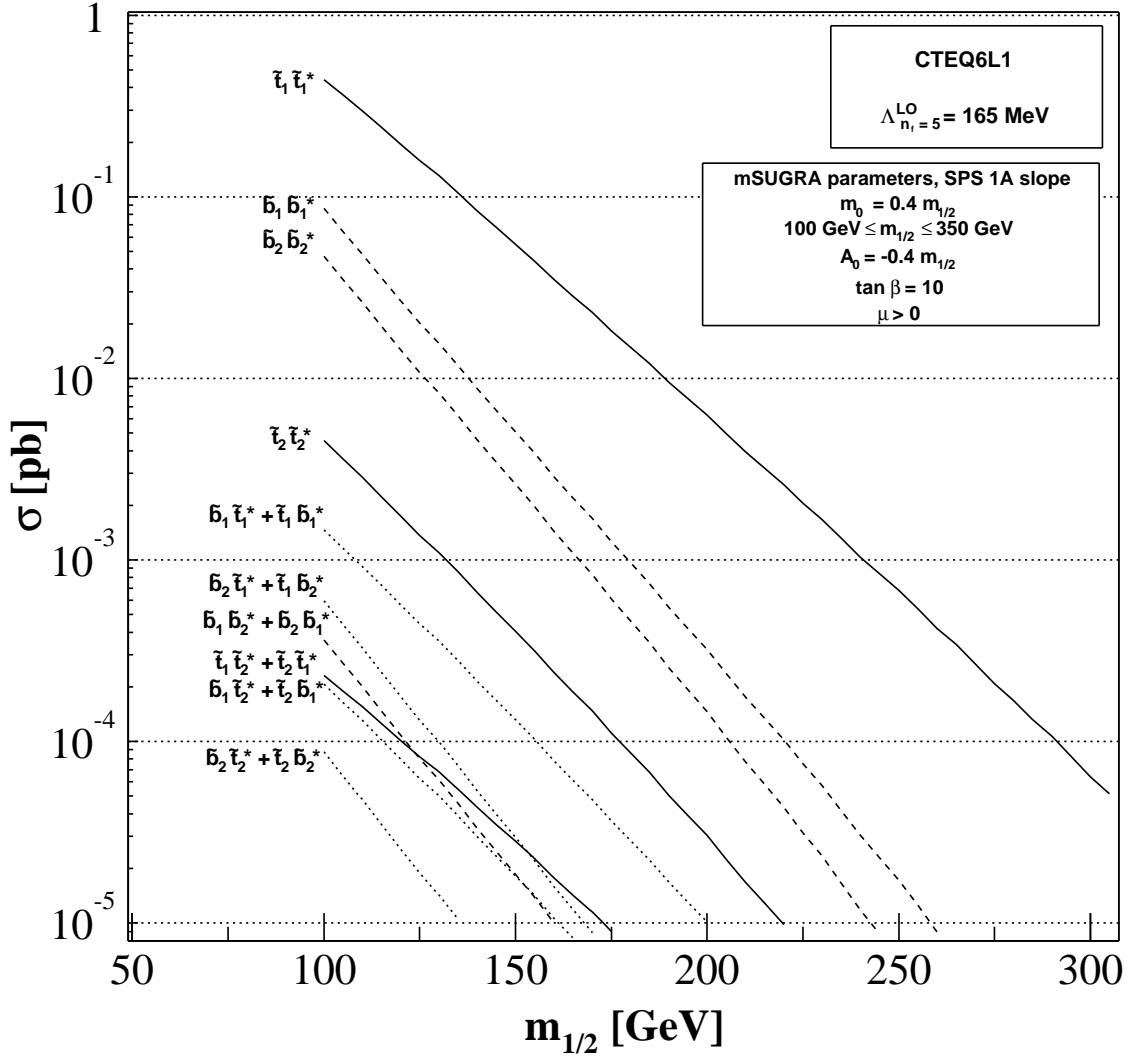}
 \caption{\label{fig:6}Production cross sections for top (full), bottom
 (dashed), and mixed top and bottom squarks (dotted) at the Tevatron as a
 function of the common fermion mass $m_{1/2}$ in the mSUGRA model line SPS
 1a \cite{Allanach:2002nj}.}
\end{figure}

\begin{figure}
 \centering
 \includegraphics[width=\columnwidth]{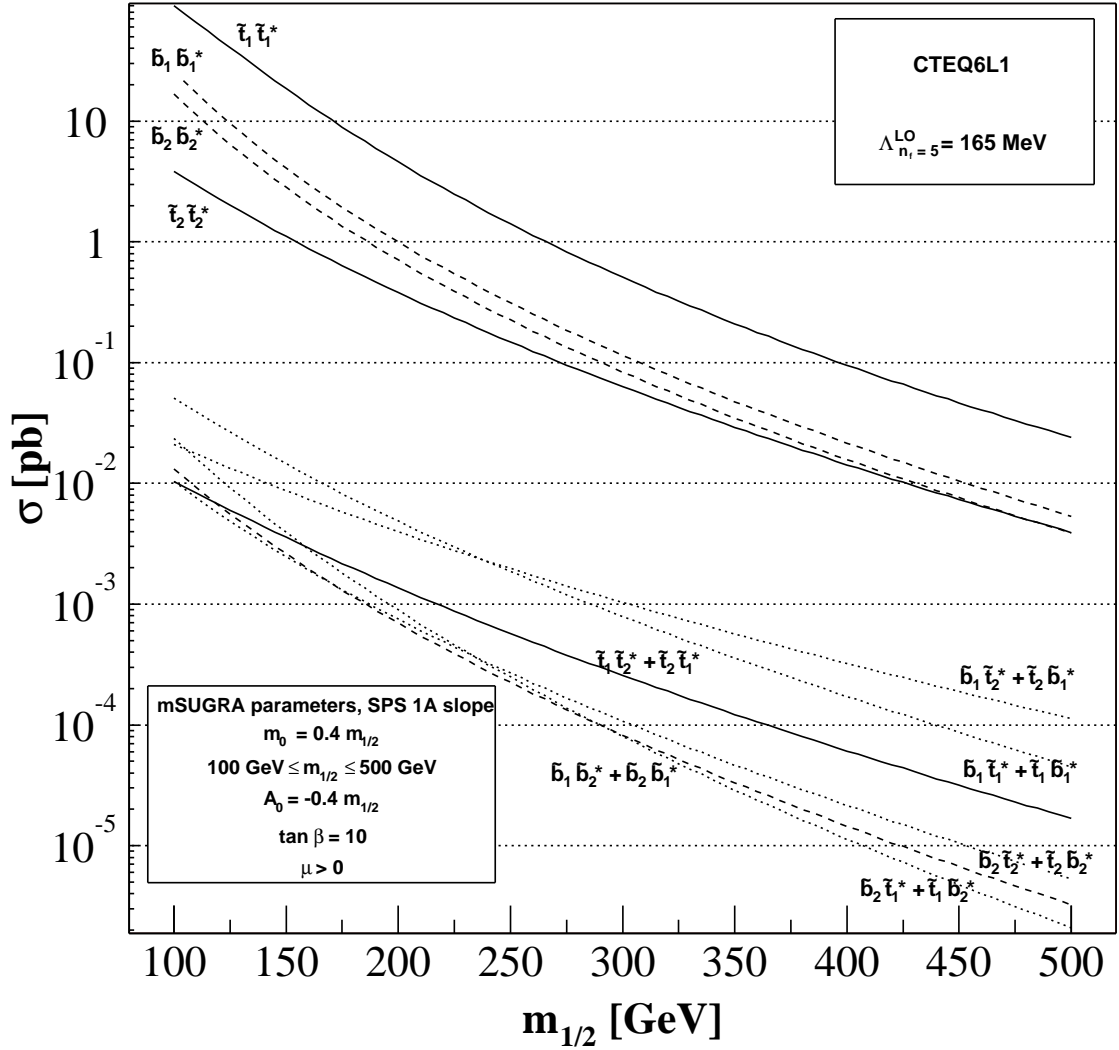}
 \caption{\label{fig:7}Production cross sections for top (full), bottom
 (dashed), and mixed top and bottom squarks (dotted) at the LHC as a
 function of the common fermion mass $m_{1/2}$ in the mSUGRA model line SPS
 1a \cite{Allanach:2002nj}.}
\end{figure}

SPS 1a is a typical mSUGRA point with an intermediate value of $\tan\beta=
10$ and $\mu>0$. It has a model line attached to it, which is specified by
$m_0=-A_0=0.4~m_{1/2}$. We vary the common fermion mass $m_{1/2}$ from 100
GeV, where $m_{\tilde{t}_1}=177$ GeV and $m_{\tilde{b}_1}=229$ GeV lie
already considerably above the current exclusion limits of 95.7 and 89 GeV
\cite{Eidelman:2004wy}, to 300 GeV for the Tevatron (Fig.\ \ref{fig:6}) and
and 500 GeV for the LHC (Fig.\ \ref{fig:7}), respectively. At the benchmark
point, $m_{1/2}=250$ GeV, leading to relatively heavy $\tilde{t}_1$ and
$\tilde{b}_1$ masses of 399 and 521 GeV.

With the limited Tevatron center-of-mass energy of $\sqrt{S}=1.96$ TeV,
non-diagonal and mixed squark production will be difficult to discover. As
one can see in Fig.\ \ref{fig:6}, only diagonal production of the lighter
%
%
top squark mass eigenstate will be visible in the full region of the mSUGRA
parameter space shown here with the expected final integrated luminosity of
8.9 fb$^{-1}$. For diagonal sbottom production, the accessible parameter
space is already reduced to $m_{1/2}\leq 225$ GeV. Non-diagonal and mixed
squark production could only be discovered, if the common fermion mass
$m_{1/2}$ is not much larger than 100 GeV. As expected for a $p\bar{p}$
collider, the cross sections are very much dominated by $q\bar{q}$
annihilation, even for the diagonal channels, and $gg$ initial states
contribute at most 15\% in the case of diagonal light stop production.

The LHC with its much larger center-of-mass energy of $\sqrt{S}=14$ TeV
and design luminosity of 300 fb$^{-1}$ will, in contrast, have no problem in
producing all combinations of squarks in sufficient numbers. The hierarchy
between the strong diagonal production channels of ${\cal O}(\alpha_s^2)$
and the electroweak non-diagonal and mixed channels of ${\cal O}(\alpha^2)$
is, however, clearly visible in Fig.\ \ref{fig:7}, the latter being about
two orders of magnitude smaller.
%
%
The LHC being a $pp$ collider and the average $x$-value in the PDFs being
considerably smaller, the diagonal channels are enhanced by the high $gg$
luminosity, which dominates their cross sections by up to 93 \%. Among the
electroweak ${\cal O}(\alpha^2)$ processes, mixed production of top and
bottom squarks is favored over non-diagonal top or bottom squark production
by the possibility of two light masses and a positive charge in the final
state, which is more easily produced by the charged $pp$ initial state.

\begin{figure}
 \centering
 \includegraphics[width=\columnwidth]{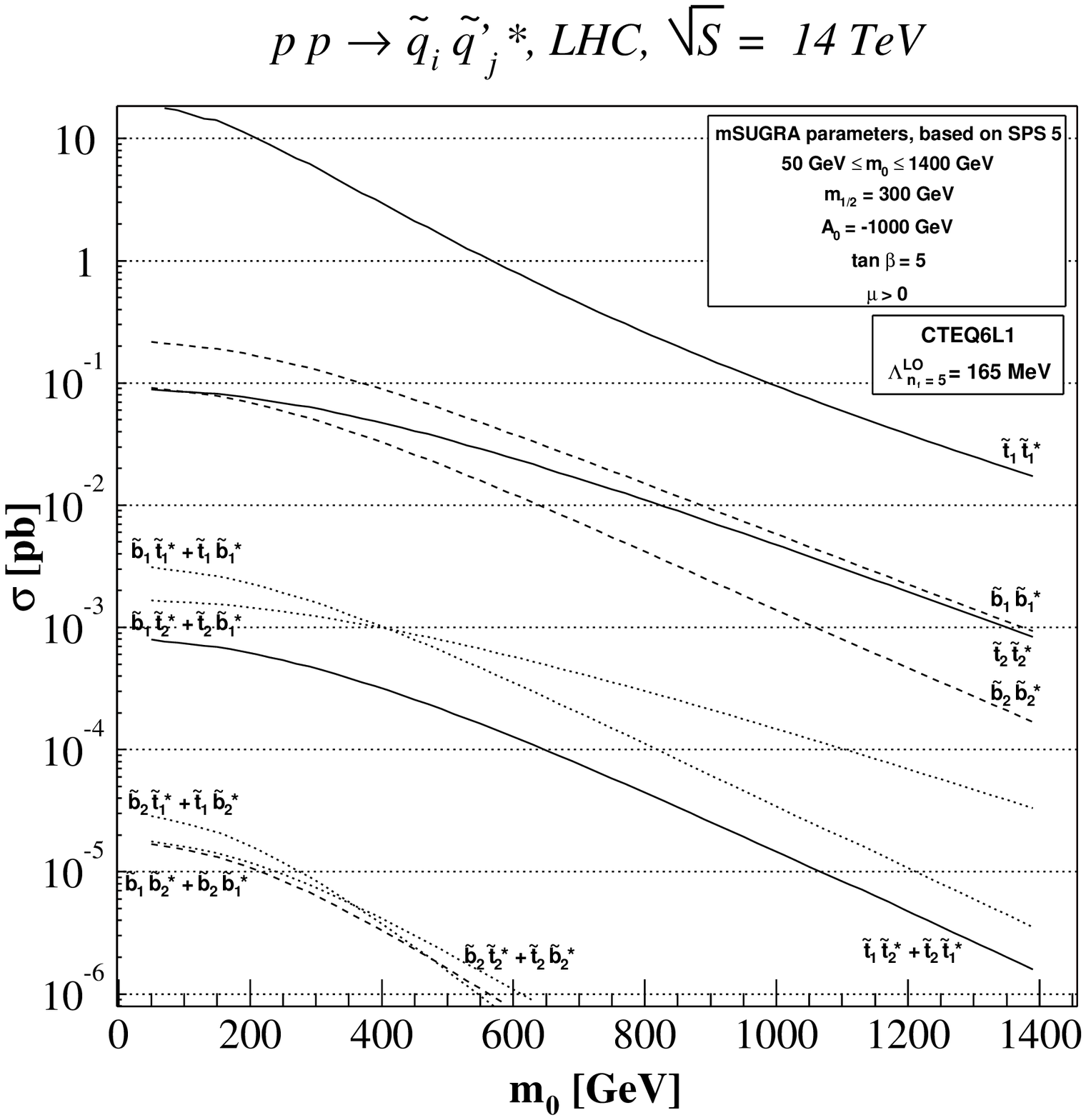}
 \caption{\label{fig:8}Production cross sections for top (full), bottom
 (dashed), and mixed top and bottom squarks (dotted) at the LHC as a
 function of the scalar mass $m_0$ in the mSUGRA model SPS 5
 \cite{Allanach:2002nj} with a light top squark.}
\end{figure}

\begin{figure}
 \centering
 \includegraphics[width=\columnwidth]{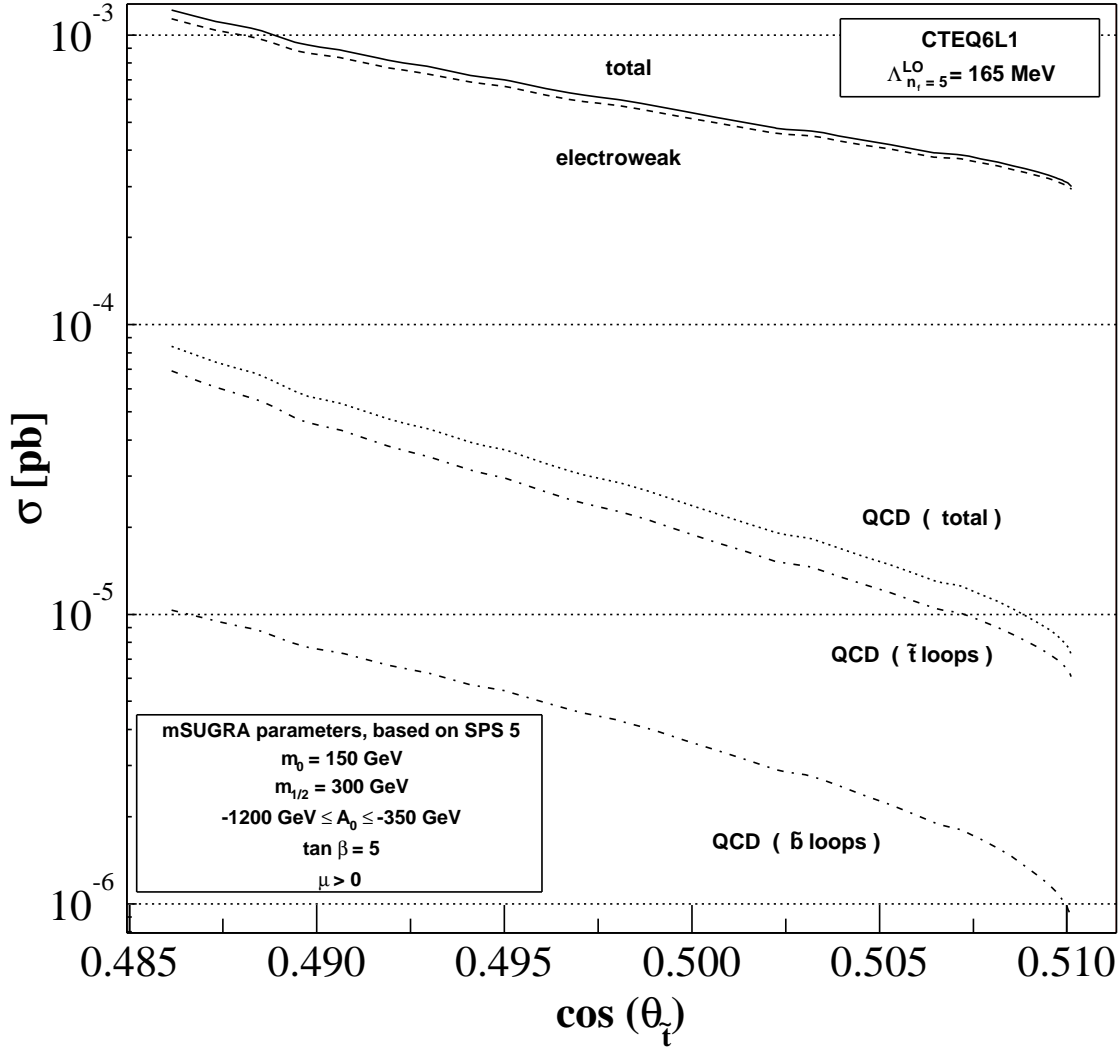}
 \caption{\label{fig:9}Contributions of tree-level electroweak and
 loop-level QCD processes to non-diagonal stop production at the LHC in
 the mSUGRA model SPS 5 \cite{Allanach:2002nj} with a light top squark as a
 function of the cosine of the top squark mixing angle, together with their
 sums.}
\end{figure}

SPS 5 is a slightly different mSUGRA scenario with lower $\tan\beta=5$,
larger $m_{1/2}=350$ GeV, and large negative $A_0=-1000$ GeV, leading to
heavier sbottoms of 566 and 655 GeV, a heavy $\tilde{t}_2$ of 651 GeV, but
also a light $\tilde{t}_1$ of 259 GeV. Since non-diagonal and mixed squark
production will be unaccessible in this case at the Tevatron, we show in
Figs.\ \ref{fig:8} and \ref{fig:9} numerical results for the LHC only,
varying either $m_0$ (Fig.\ \ref{fig:8}) or $A_0$ (Fig.\ \ref{fig:9})
independently to test the sensitivity of the cross section on the squark
masses and mixing.

Concentrating first on the $m_0$-dependence, we see in Fig.\ \ref{fig:8}
%
%
a clear hierarchy between the dominant pair production of the lighter stop,
strong pair production of the heavier stop and sbottoms, charged and neutral
electroweak production of final states involving at least one light squark,
and finally charged and neutral electroweak production of the heavier
squarks, which may only be visible up to $m_0\leq 600$ GeV. The more
pronounced hiearchy in Fig.\ \ref{fig:8} can be explained by the
considerable squark mass differences in SPS 5, which lead to additional
phase space suppression for the heavier squarks.

As mentioned in the Introduction, the case of non-diagonal squark production
merits a more detailed investigation of the relative importance of the
coupling-suppressed electroweak diagram in Fig.\ \ref{fig:2} and the
loop-suppressed QCD diagrams in Fig.\ \ref{fig:4}. As mentioned in Sec.\
\ref{sec:2}, the $t$-channel gluino and neutralino exchange diagrams in
Fig.\ \ref{fig:3} are only present for sbottom production. For the mSUGRA
scenario considered here, their contributions are found to be six to eight
orders of magnitude smaller than the $s$-channel contribution in Fig.\
\ref{fig:2}. While one naively expects the ${\cal O}(\alpha^2)$ and
${\cal O}(\alpha_s^4)$ diagrams in Figs.\ \ref{fig:2} and \ref{fig:4} to
contribute with roughly equal strength, Fig.\ \ref{fig:9} shows for the
%
%
case of non-diagonal stop production in the SPS 5 scenario and at the LHC
that the QCD loop-contributions are smaller than the electroweak tree-level
contribution by about one order of magnitude. This is easily explained by
the presence of additional heavy squark propagators in the loop diagrams.
Here, we consider not only loops involving top, but also bottom squarks,
which do not cancel in Eq.\ (\ref{eq:14}), if the masses of the two sbottom
mass eigenstates are unequal. However, the non-diagonal elements in the
squark mass matrices are proportional to the relevant SM quark mass and
$m_b\ll m_t$, so that mixing effects are less important for sbottoms than
for stops. Consequently, sbottom loops contribute about one order of
magnitude less than stop loops, as can also be seen in Fig.\ \ref{fig:9}.
SUSY-QCD loop diagrams involving gluino exchanges have not been calculated
here, as they are of ${\cal O}(\alpha_s^4)$ and require in addition the
presence of heavy top quark and gluino propagators in the loop. In the SPS 5
scenario, we have indeed a heavy gluino of mass $\mg=725$ GeV.
The tree-level cross section in Eq.\ (\ref{eq:7}) and the sbottom loop
contribution in Eq.\ (\ref{eq:14}) depend on the cosine of the stop mixing
angle through $\sin^2(2\theta_{\tilde{t}})=4\cos^2\theta_{\tilde{t}}~(1-
\cos^2\theta_{\tilde{t}})$, which is clearly visible in Fig.\ \ref{fig:9}.
In contrast, the stop loop contribution in Eq.\ (\ref{eq:14}) has a steeper
dependence through $\sin^2(4\theta_{\tilde{t}})=16\cos^2\theta_{\tilde{t}}~
(1-\cos^2\theta_{\tilde{t}})~(1-2\cos^2\theta_{\tilde{t}})^2$, which is
also visible in Fig.\ \ref{fig:9}.

\begin{figure}
 \centering
 \includegraphics[width=\columnwidth]{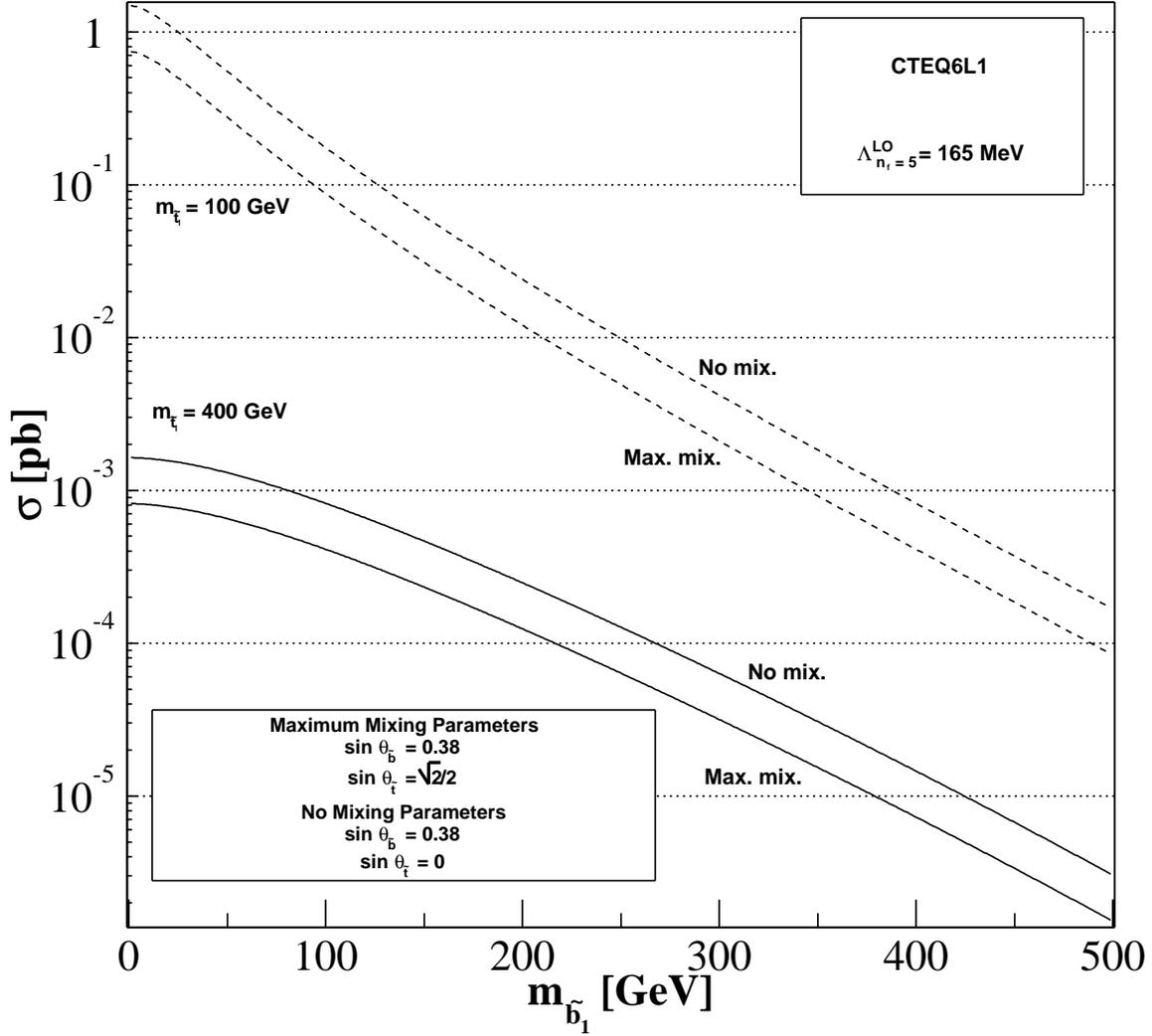}
 \caption{\label{fig:10}Associated production of top and bottom squarks at
 the Tevatron as a function of a light sbottom mass \cite{Berger:2000mp} for
 $m_{\tilde{t}_1}~=~100$ GeV (dashed) and 400 GeV (full) and no (top) or
 maximal (bottom) top squark mixing.}
\end{figure}

The apparent excess of the experimentally observed bottom cross section
at the Tevatron over the QCD prediction has led to speculations of light
sbottom contributions to this cross section \cite{Berger:2000mp}, although
the use of up-to-date information on the $B$ fragmentation function may
reduce the discrepancy to an acceptable level \cite{Cacciari:2002pa}. Taking
the light sbottom scenario at face-value with a sbottom mixing angle of
$\sin \theta_{\tilde{b}}=0.38$ that reduces its coupling to the $Z$-, but
not the $W$-boson, we show in Fig.\ \ref{fig:10} that mixed
%
%
production of light top and bottom squark mass eigenstates at the Tevatron
is yet another promising channel to confirm or exclude this scenario, as
the cross section is well visible over a large range of $\tilde{b}_1$
masses, almost independently of the stop mixing angle and for light
($m_{\tilde{t}_1}=100$ GeV) as well as for heavier ($m_{\tilde{t}_1}=400$
GeV) top squarks.

\begin{figure}
 \centering
 \includegraphics[width=\columnwidth]{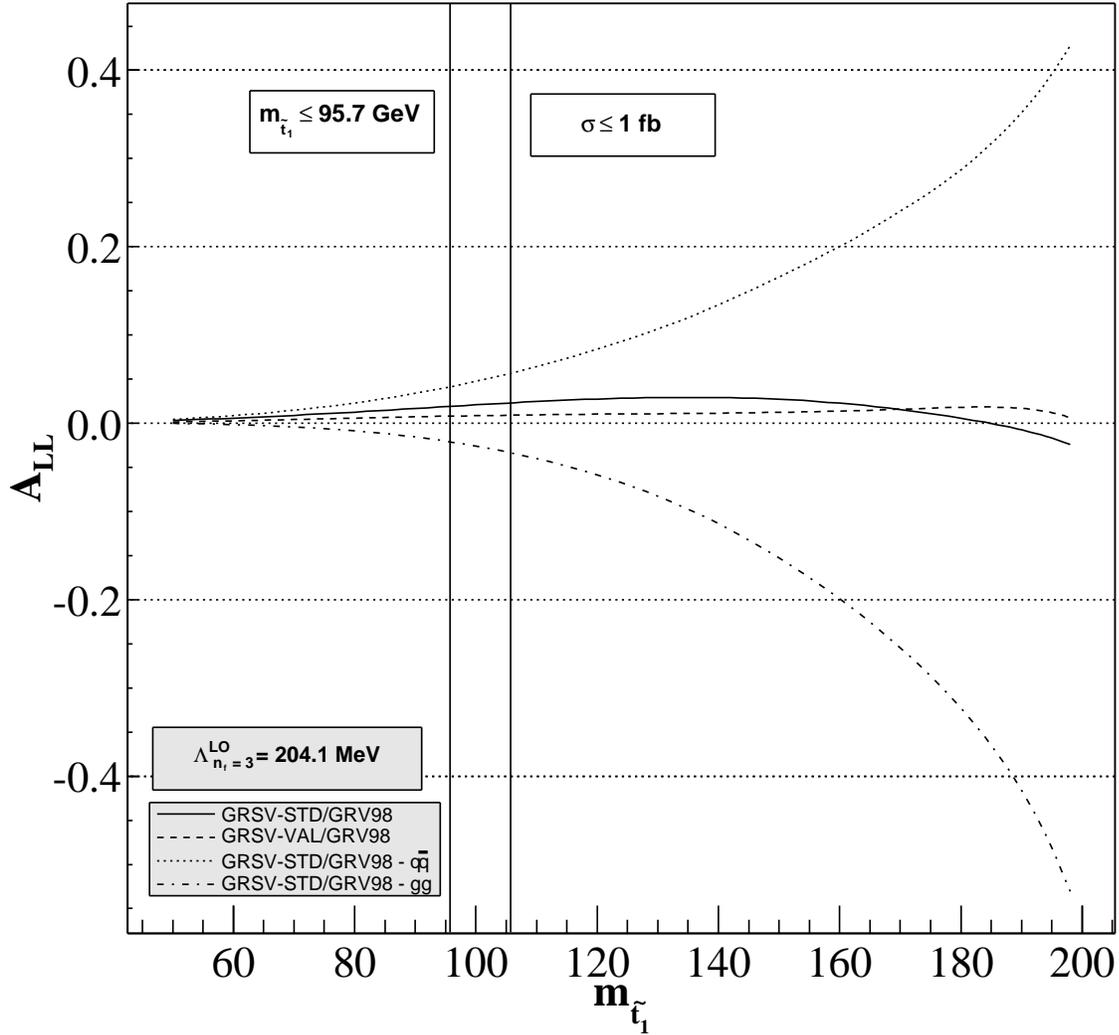}
 \caption{\label{fig:11}Contributions of the $q\bar{q}$ (dotted) and $gg$
 (dot-dashed) initial states to the longitudinal double-spin asymmetry
 $A_{LL}$ at RHIC, together with their sum (full), for stop pair production
 as a function of the light top squark mass. The total asymmetry using
 GRSV-VAL (dashed) instead of GRSV-STD (full) parton densities
 \cite{Gluck:2000dy} is also shown.}
\end{figure}

As a possible application of the polarization dependence of our analytical
results, we show finally in Fig.\ \ref{fig:11} the double-spin asymmetry
%
%
of diagonal light stop production at RHIC. As this polarized $pp$ collider
has only a rather small center-of-mass energy of $\sqrt{S}=500$ GeV, the
observable stop mass range is obviously very limited. Already for
$m_{\tilde{t}_1}>106$ GeV, the unpolarized cross section drops below 1 fb,
while stop masses below 96 GeV are most likely already excluded
\cite{Eidelman:2004wy}. This leaves only a very small mass window of 10 GeV
for possible observations. In Fig.\ \ref{fig:11} one can clearly see the
rise of the asymmetry for $q\bar{q}$ and $gg$ initial states, as the stop
mass and the correlated $x$-value in the PDFs grows. However, as the
two asymmetries are approximately of equal size, but opposite sign, the
total observable asymmetry rests below the 5\% level in the entire mass
range shown. This is true for both choices of polarized parton densities,
GRSV 2000 standard (STD) as well as valence (VAL) \cite{Gluck:2000dy}.
For consistency, the unpolarized cross sections have been calculated in this
case using the GRV 98 parton density set \cite{Gluck:1998xa}.

\section{Conclusion}
\label{sec:4}

In Summary, we have presented in this Paper the most extensive analysis to
date of diagonal, non-diagonal, and mixed squark production at hadron
colliders. Great care has been taken to include in all cases the dominant
contributions mediated not only by strong, but also by electroweak
interactions up to ${\cal O}(\alpha_s^4)$ and ${\cal O}(\alpha^2)$,
respectively. Squared helicity amplitudes have been presented for all
considered channels in analytic form, as they expose the left- and
right-handed contributions in the electroweak channels, allow for future
applications to polarized hadron collisions, and may easily be implemented
in general purpose Monte Carlo programs.

Numerically, we have focused on top and bottom squark production, including
mixing in both cases, at the Tevatron and the LHC. We have emphasized the
fact that associated light sbottom and stop production may allow for
confirmation or exclusion of light sbottom scenarios at the Tevatron. In
more traditional scenarios such as the SPS 1a or SPS 5 models, non-diagonal
and mixed squark production can probably only be studied at the LHC, where
these channels may allow for additional constraints on SUSY masses, mixing
angles, or the SUSY CKM matrix. \\

\noindent {\em Note added in proof:} After presentation of our work at the
GDR SUSY conference in Grenoble \cite{grenoble} and at the Cortona 2005
Theoretical Physics Meeting \cite{cortona}, and while this Paper
was being completed, a second publication related to the mixed top and
bottom squark production aspects of our paper appeared
\cite{Berdine:2005tz}. Unfortunately, the authors do not present any cross
section formul\ae, so that analytical comparisons are impossible. However,
we have checked that our calculations agree with their numerical $s$-channel
results in Tab.\ I. Note that real emission contributions such as those
labeled $t$-channel in Tab.\ I, coming from gluon initial states, cannot be
reliably calculated separately, as they partially contain initial-state
singularities that must be absorbed in the proton PDFs, so that these
contributions must be included in a complete next-to-leading order
calculation.

\section*{Acknowledgments}
We thank S.\ Calvet and T.\ Millet for valuable discussions at the GDR SUSY
conference in Grenoble. This work was supported by a postdoctoral and a
Ph.D.\ fellowship of the French ministry for education and research.
%

\appendix
\section{Squark Mixing}
\label{sec:a}

The (generally complex) soft SUSY-breaking terms $A_q$ of the trilinear
Higgs-squark-squark interaction and the (also generally complex)
off-diagonal Higgs mass parameter $\mu$ in the MSSM Lagrangian induce
mixing of the left- and right-handed squark eigenstates $\tilde{q}_{L,R}$ of
the electroweak interaction into mass eigenstates $\tilde{q}_{1,2}$. The
squark mass matrix \cite{Haber:1984rc,Gunion:1984yn}
\beq
 {\cal M}^2 =
 \lr\begin{array}{cc}
  m_{LL}^2+m_q^2  &
  m_q m_{LR}^\ast \\
  m_q m_{LR}      &
  m_{RR}^2+m_q^2
 \end{array}\rr
\eeq
with
\bea
 m_{LL}^2&=&(T_q^3-e_q\sin^2\theta_W)m_Z^2\cos2\beta+m_{\tilde{Q}}^2,\\
 m_{RR}^2&=&e_q\sin^2\theta_W m_Z^2\cos2\beta+\left\{\begin{array}{l}
 m_{\tilde{U}}^2\hspace*{4.8mm}{\rm for~up-type~squarks},\\
 m_{\tilde{D}}^2\hspace*{4.5mm}{\rm for~down-type~squarks},\end{array}\right.\\
 m_{LR}  &=&A_q-\mu^\ast\left\{\begin{array}{l}
 \cot\beta\hspace*{5mm}{\rm for~up-type~squarks}\\
 \tan\beta\hspace*{4.5mm}{\rm for~down-type~squarks}\end{array}\right.
\eea
is diagonalized by a unitary matrix $S$, $S {\cal M}^2 S^\dagger={\rm diag}\,
(m_1^2,m_2^2)$, and has the squared mass eigenvalues
\beq
 m_{1,2}^2=m_q^2+{1\over 2}\lr m_{LL}^2+m_{RR}^2\mp\sqrt{(m_{LL}^2-m_{RR}^2)^2
 +4 m_q^2 |m_{LR}|^2}\rr.
\eeq
For real values of $m_{LR}$, the squark mixing angle $\theta_{\tilde{q}}$,
$0\leq\theta_{\tilde{q}}\leq\pi/2$, in
\beq
 S = \lr \begin{array}{cc}~~\,\cos\theta_{\tilde{q}} &
                              \sin\theta_{\tilde{q}} \\
                             -\sin\theta_{\tilde{q}} &
                              \cos\theta_{\tilde{q}} \end{array} \rr
 \hspace*{3mm} {\rm with} \hspace*{3mm}
   \lr \begin{array}{c} \tilde{q}_1 \\ \tilde{q}_2 \end{array} \rr =
 S \lr \begin{array}{c} \tilde{q}_L \\ \tilde{q}_R \end{array} \rr
\eeq
can be obtained from
\beq
 \tan2\theta_{\tilde{q}}={2m_qm_{LR}\over m_{LL}^2-m_{RR}^2}.
\eeq
If $m_{LR}$ is complex, one may first choose a suitable phase rotation
$\tilde{q}_R'=e^{i\phi}\tilde{q}_R$ to make the mass matrix real and then
diagonalize it for $\tilde{q}_L$ and $\tilde{q}_R'$. $\tan\beta=v_u/v_d$
is the (real) ratio of the vacuum expectation values of the two Higgs fields,
which couple to the up- and down-type (s)quarks. The weak isospin quantum
numbers for left-handed up- and down-type (s)quarks with hypercharge $Y_q=1/3$
are $T_q^3=\{+1/2,-1/2\}$, whereas $Y_q=\{4/3,-2/3\}$ and $T_q^3=0$ for
right-handed (s)quarks, and their fractional electromagnetic charges are
$e_q=T_q^3+Y_q/2$.
The soft SUSY-breaking mass terms for left- and right-handed squarks
are $m_{\tilde{Q}}$ and $m_{\tilde{U}}$, $m_{\tilde{D}}$, respectively,
and $m_Z$ is the mass of the neutral electroweak gauge boson $Z^0$.



\end{document}